**Critical review**

A materials perspective on the design of damage-resilient artificial bones and bone implants through additive/advanced manufacturing.

**Authors**

Hortense Le Ferrand[1,*], Christos E Athanasiou[2]

**Affiliations**

[1] School of Mechanical and Aerospace Engineering, School of Materials Science and Engineering, Nanyang Technological University, Singapore

[2] School of Engineering, Brown University, Providence, USA

**Corresponding author:**

*Hortense@ntu.edu.sg

**Abstract**

After more than five decades of research, the failure of bone implants is still an issue that becomes increasingly urgent to solve in our ageing population. Among the reasons for failure, catastrophic brittle fracture is one event that is directly related to the implant's material and fabrication and that deserves more attention. Indeed, clinically available implants pale at reproducing the hierarchical and heterogeneous microstructural organization of our natural bones, ultimately failing at reproducing their mechanical strength and toughness. Nevertheless, the recent advances in additive and advanced manufacturing open new horizons for the fabrication of biomimetic bone implants, challenging at the same time their characterization, testing and modelling. This critical review covers selected recent achievements in bone implant research from a materials standpoint and aims at deciphering some of the most urgent issues in this multidisciplinary field.

## 1. Introduction: Failure of implants

With ageing, our skeleton weakens from the combined effect of osteoporosis (bone loss), sarcopenia (muscle loss), dehydration, diseases. Joints that are submitted to large cycling stresses are particularly sensitive. As a consequence, hip and knee implants are prevalent among the elderly and were estimated to concern 5.26% and 10% of the US population above 80 years old in 2010 [1]. These data are predicted to increase by 2 to 3% by 2050, with most patients now living in Asia [2]. But bone implants are not the appanage of the elderly only as sports injuries, diabetes or unhealthy nutrition may also lead to the need for bone implantation at a younger age. Although the current implants are generally satisfactory for patients above 80, they pose problems for the younger generation. The risk of having a hip implant revision and a second surgery is estimated at 10% for patients aged above 75, and up to 20% for patients below 60. These numbers are 5% higher for knee replacements [3,4] and represent significant costs for the patients and the health care systems [5,6].

There may be numerous reasons for an implant to fail: a fault during the surgical operation, an infection, allergy or disease, rejection from the immune system, inflammation, unhealthy lifestyle, an accident. In this review, only failures related to the material aspect of the implant will be discussed, namely its composition, organization and design. These failures, thus, correspond either to the fracture of the implant material itself, either to the fracture of the bone underneath the implant due to mismatching mechanical properties. First, the general context of implant failures at the material's level and research in this area are described. Second, materials selection and organization to achieve bone-like mechanical properties are detailed and the limitations of the strategies involved discussed. Third, the most recent efforts in additive manufacturing towards long-term damage resistant implants are reviewed and their contribution to more appropriate bone-like designs and mechanics highlighted. Finally, the methods employed to characterize those new bone-like implants are reported and suggestions are made on how to align laboratory experiments with clinical tests for a more representative assessment of their performance.

## 1.1. Literature overview

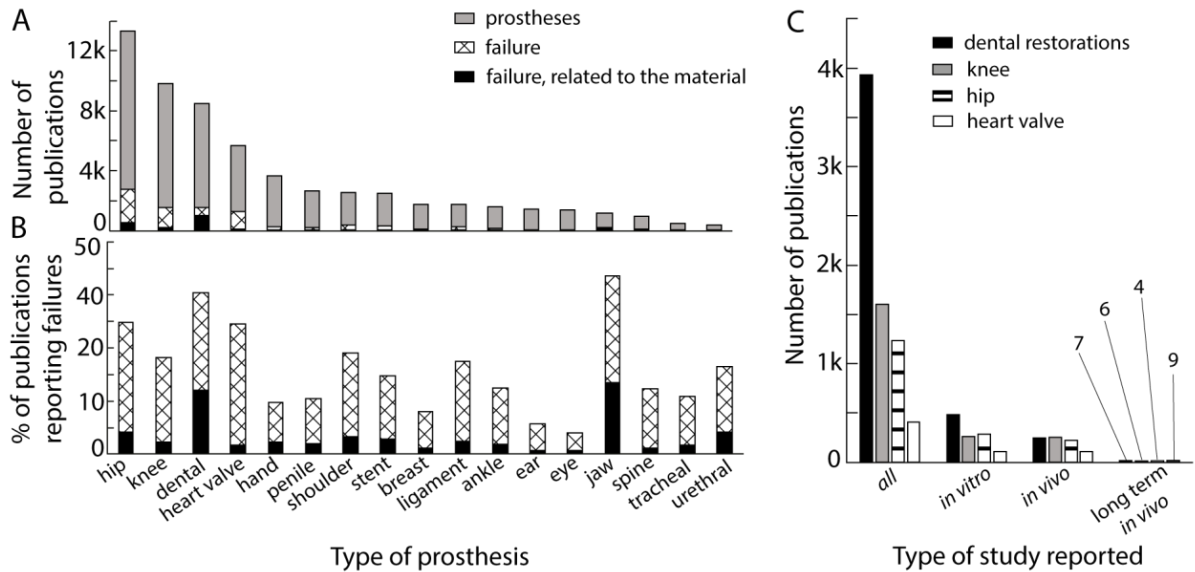

**Figure 1: Overview of the literature on prostheses and material related failure. A)** Number of publications as a function of the type of prosthesis, obtained through Isi Web of Knowledge using the following keywords: "prostheses" (grey), "prostheses" and "failure" (black crosses) and "protheses" and "failure" and "materials" (black), from 1900 till 2019. **B)** Percentage of total publications reporting failures as a function of the type of prosthesis, using the similar legend. **C)** Number of publications as a function of the type of study carried out for dental restorations (black), knee implants (grey), hip joints (striated) and heart valves (white).

There is an extensive literature on the fabrication and characterization of prosthetic materials, covering a large range of tissues **(Figure 1A)**. Although prosthetic materials and implants might be differentiated in some dictionaries, they are often referring to the same devices. Therefore, we consider them as synonyms in this review (Table 1 in the Electronic Supplementary Material (ESM)). Since 1900, Isi Web of Knowledge counts a total of ~ 60 k publications on protheses, among which ~ 9.5 k, *i.e*. 15%, report failure. Among those 9.5 k, ~ 26% specify that the failure is related to the material, *i.e.* 4% of the total set of publications. Hard tissues and joints have a privileged focus with already more than 30 k publications in total and more than 500 papers published every year since 2010. The second focus of implant research is on dental restorations with a smooth increase from ~ 430 papers in 2010 to ~ 800 in 2018. Despite such a large research output, these hard implants are still prone to dramatic failure. 20 to 45% of the total publication set are reporting failures of these implants and up to 13% of those failures are related to their material composition and design only **(Figure 1B).** Over the range of implants screened in the literature, the number of papers reporting failures at the materials level is the highest for jaw prostheses. This is probably due to the geometrical complexity of the jaw, comprising anchoring for teeth and the two joints at the mandibula. In addition, jaws undergo continuous cycling loading in changing liquid environment during biting, chewing, or talking. Although screening the literature is informative, it is likely that the occurrence of failure in real clinical cases overcomes the number of studies on it, but it already denotes the need for improvement of the materials structure of bone implants.

Along with innovations on prosthetic materials, there is also a need to develop the adapted methods to measure, understand and predict their long-term performance. This is critical as most material-related bone implant fractures occur in patients only after long-term implantation and loading history, *i.e.* after 10 to 15 years. This clinical observation raises a large number of questions, such as: How to reproduce 10-15 years of variable dynamic loadings and chemical variations in a short-term laboratory experiment? What *in vitro* or *in vivo* models to use? What loading model to apply? How representative of the human body are those models? Can we have "one implant fits all patients"? These questions remain largely unanswered and the number of long-term *in vivo* and clinical studies is still too little (**Figure 1C** and Tables 2-5 in the ESM). Such a low publication number highlights the additional difficulties to reach the stage of long-term *in vivo* stage, to share the data and to assess the causes of the failure.

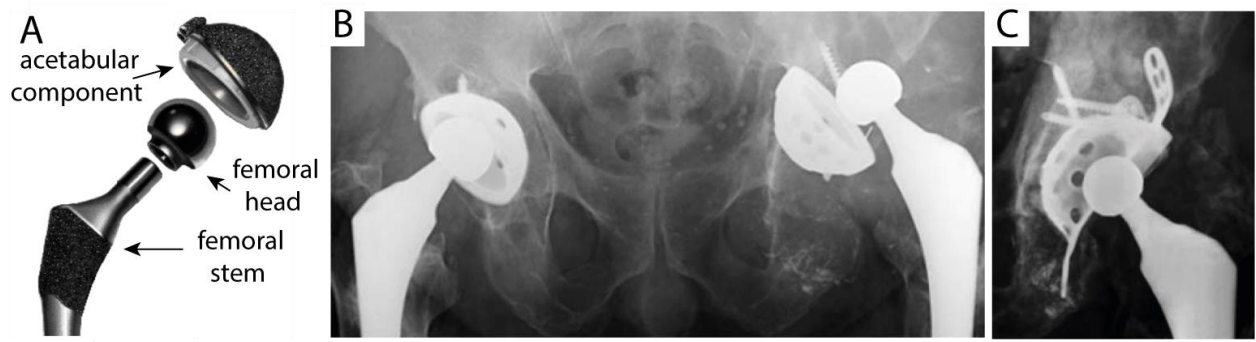

**Figure 2: Clinical case of a hip joint long-term failure.** A) Image of a clinically available hip joint implant (adapted from [7], copyright © 2009, Scientific.Net). B) Radiography of the patient 16 years after first surgery showing the displacement of the acetabular component of the implant due to bone resorption (osteolysis) on the left leg only and (C) radiography of this implant after the revision ([8], copyright © 2008, Journal of the American Academy of Orthopaedic Surgeon).

As an illustration to such long-term implant failures, Marshall *et al*. reported the case of a 72-years old man who did not return to his post-operation follow-up session after a double hip joint replacement (**Figure 2**). The patient, after 16 years, developed pain and leg shortening on the left side [8]. Using X-ray scans, it was revealed that his left pelvic bone severely resorbed, rotating the acetabular part of the hip joint implant that is anchored in the pelvis and that serves as a socket to the hip prosthesis (**Figure 2A**). As a consequence of this motion, the bottom part of the implant, the femoral stem, moved up, pulling along the left leg up. On the right side, no resorption, also termed osteolysis, had occurred and the implant and the length of the leg remained the same. This case illustrates the complexity of the problem: why did bone resorption occur and why only on one side? As a "silent" disease whose progression is only revealed after 5 to 10 years, osteolysis is challenging the assessment of the long-term performance of implants. To complexify this further, the lifestyle, health and metabolism of the patient are also key factors impacting the performance of the implant. For example, an aged person is likely to develop osteoporosis and increased bone resorption; a patient involved in demanding sports might submit the implant to unusual stresses, whereas a patient with a past injury might also have weak points in the bones.

### 1.2. Causes of materials failure

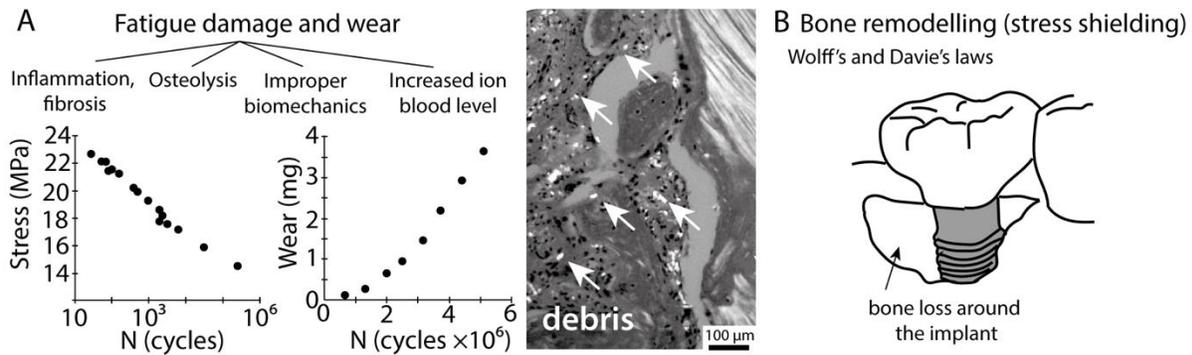

**Figure 3: Main causes of material-related failures of implants:** A) consequences of fatigue damage and wear. The plots represent the stress required for failure and the wear as a function of the number of cycles for UHMWPE (data extracted from [9,10]). The histology image shows debris particles (white arrows) from a UHMWPE implant and the fibrous tissue formation with multinucleated giant cells having taken up the particles *via* endocytosis (adapted from [11], copyright © 2007, Wolters Kluwer Health, Inc.); B) bone remodeling around the implant material

due to Wolff's and Davie's laws for implants with mechanical properties that do not match the underlying bone.

The predominant cause of implant failure when related to its material properties is due to fatigue damage and wear during cyclic loading **(Figure 3A)** [12]. Resistance to fatigue crack propagation is the intrinsic property of a material to sustain cracks that might develop during loading. A material will low fatigue resistance can sustain stresses of decreasing intensities as the number of cycle increases (left plot in **Figure 3A**). As a first order approximation this is inversely proportional to the wear (right plot of **Figure 3A**). A material with low fatigue resistance, *i.e.* a brittle material, will catastrophically fail at the first crack. Damage-resistant materials, on the other hand, can still withstand a certain amount of loading despite the presence of multiple cracks [13]. Such materials are said to be either tough, meaning that it costs a lot of energy to initiate cracking, or to have toughening mechanisms in their structure to resist against crack propagation [14]. The other mechanism that leads to failure is wear. Wear results from micromotions between the implant and the anchoring bone, or between the different parts of a joint prothesis. In the body, wear is significantly reduced by a cartilage layer soaked with a lubricant, the synovial fluid. Cartilage is a highly hydrated tissue that despite containing up to 80 wt% of water [15], exhibits high strength of about 35 MPa [16] and toughness of ~11 $GPa\sqrt{m}$ [17]. To avoid the hard-to-hard contacts between two implants at joints, ceramic or polymers have been preferred to metals. Nevertheless, the micro- and nanoparticles worn off during loading cause inflammation and fibrosis in the surrounding tissue [18,19]. To avoid bone to implant friction, fixation and stability is usually ensured using bone cements and bioactive coatings that enhance bone growth at the surface of the implant. In addition to the formation of nanoparticles, the wear of implant materials can increase the ion content in body fluids [20]. Finally, wear is further worsened by oxidation and the changes in temperature or in pH that occur in the body [21,22].

Along with fatigue damage and wear, the other main cause of failure is related to bone remodeling and stress shielding underneath the implant **(Figure 3B)**. Indeed, implants with mechanical properties overpassing that from the anchoring bone will take up most of the load. This leads to stress shielding, which means that the load will not be transmitted to the bone underneath the implant. As a result, the bone part that is not being challenged mechanically will decay. This phenomenon, referred to as the Wolff's (for hard tissues) or the Davie's law (for softer tissues) [23] is due to the mechanosensing properties of bone cells that grow bone only under mechanical stimulation [24,25]. In clinical cases, bone remodeling and stress shielding can result from an unadapted implant design or from the displacement of the implant due to micromotion or an accident. For example, Uchida *et al.* reported the case of a patient who underwent hip replacement; 20 years later, after an accident that tilted the implant from its original angle, osteolysis started and extended, requiring a second surgery 3 years later [26].

## 1.3. The gap: Mismatch between the multiscale mechanics and organization of native tissues and those of current implants.

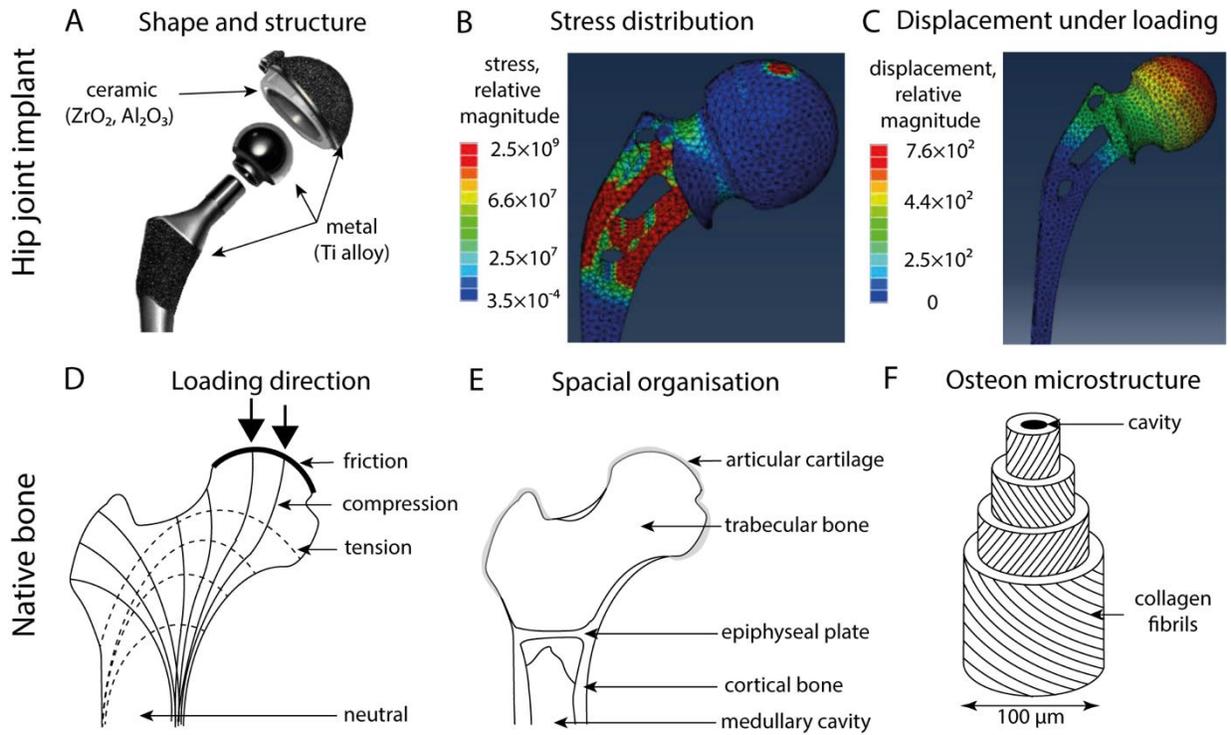

**Figure 4: Hip joint implant vs native bone: structure and mechanics.** A) Image of a typical clinically available hip joint implant (adapted from [7], copyright © 2009, Scientific.Net). Finite Element Simulations of the stress (B) and strain (C) distributions in the implant under vertical loading ([27] copyright © 2016, Elsevier Ltd.). Schematics of the stress lines in a human hip under vertical loading (D), of the meso- and macroscopic organization (E) and of the microstructure of osteons present in the cortical bone (F).

Current clinically available implants exhibit a hip-like shape, are biocompatible or bioactive, and generally stiff and strong. They are made from a handful of selected chemistries and shaped in simple geometries *via* established processes. Overall, they are a mechanically simplified version of bones. Indeed, in bones, the local nano- and microstructures, the shape, the biological response and the local mechanics are all participating in the global performance. To illustrate some of the limitations in today's implant designs, the case of a hip joint implant is discussed in the following.

A typical hip joint implant has a metallic femoral stem and head made usually in titanium alloys, whereas the acetabular component is preferably a ceramic, zirconia ($ZrO_2$) or alumina ($Al_2O_3$) (**Figure 4A**). Polymers like ultra-high-density polyethylene have been used in the past [28], whereas ceramic femoral heads can now also be found [29]. The fabrication process is well-established: the metals are forged or investment casted before being sand-blasted or polished and coated; the polymers are moulded and machined; and the ceramic parts are casted, sintered, grinded and polished [7]. The loading distribution on the implant can be predicted using solid mechanics and finite element simulations to verify it can provide and sustain vertical loading (**Figure 4B,C**) [27]. In contrast, the structure and stress distribution in native hips exhibit more heterogeneity, in particular at the micro and sub-micro levels. The effect of vertical loading at the femoral head on the stress distribution within the bone has been modelled by a mesh of stress curves that are under tension or compression (**Figure 4D**) [30]. Interestingly, in the bone design upon loading, tensile and compression stress lines develop perpendicular to each other ultimately leading to zero stress at the centre of the bone and providing damage-resistance. In addition to this, the top of the hip is covered with a layer of cartilage for wear resistance.

The possibility to combine perpendicular compensating stress directions in the same material results from its organization with local variations in density, hardness and strength (**Figure 4E**). Long bones are composed of a highly porous part, the trabecular or spongy bone whose ~80% porosity allow cell invasion and nutrient diffusion [31], and a very compact shell: the cortical or hard bone that has ~3-4 % porosity only and can take up most of the load. The cortical bone additionally features a hierarchical microstructure [32–34]. In the cortical bone, osteons are microstructural reinforcing elements and toughening elements. Osteons are highly mineralized concentric bone layers surrounding blood vessels and nerves, and are composed of mineralized collagen fibrils aligned stacked parallel to each other and tilted between each layer (**Figure 4F**) [35]. Osteons play a role in the longitudinal stiffness of bones, contributing to its anisotropy, to the deflection of cracks that might propagate perpendicularly to the long axis of the bone, and to the resistance to torsion [36–38]. At the submicrometric level, the minerals composing 65% of the walls of the osteons are made of hydroxyapatite (HA, $Ca_{10}(PO_4)_6(OH)_2$) with a needle or plate-like shape of 1-5 nm in thickness, 10-40 nm in width and 20-100 nm in length [39,40]. At the nanoscale, the collagen fibrils that hold those minerals together are formed by collagen molecules assembled in a triple helix stabilized by hydrogen bonds and exhibiting a high stiffness of 5 to 11.5 GPa [41]. Furthermore, bones are hydrated at levels varying from bone to bone and with sex, age and health [42] contributing to the changes in the mechanical properties during life [43,44]. Finally, bones show some piezoelectricity, making it an even more dynamic structure, intrinsically [45,46].

This highlights the structural and mechanical contrast between bone implants and native bones and challenges the community to fabricate materials that have more biomimetic or bioinspired structures and properties. The recent advances in materials science and manufacturing in this direction are discussed in the following.

## 2. Materials for bone-like structure and properties

To design and fabricate implants that achieve bone-like properties, there is a need to select the appropriate materials, to organise them in similar microscopic and mesoscopic structures, and this in the relevant 3 dimensional (3D) shape.

## 2.1. Materials selection

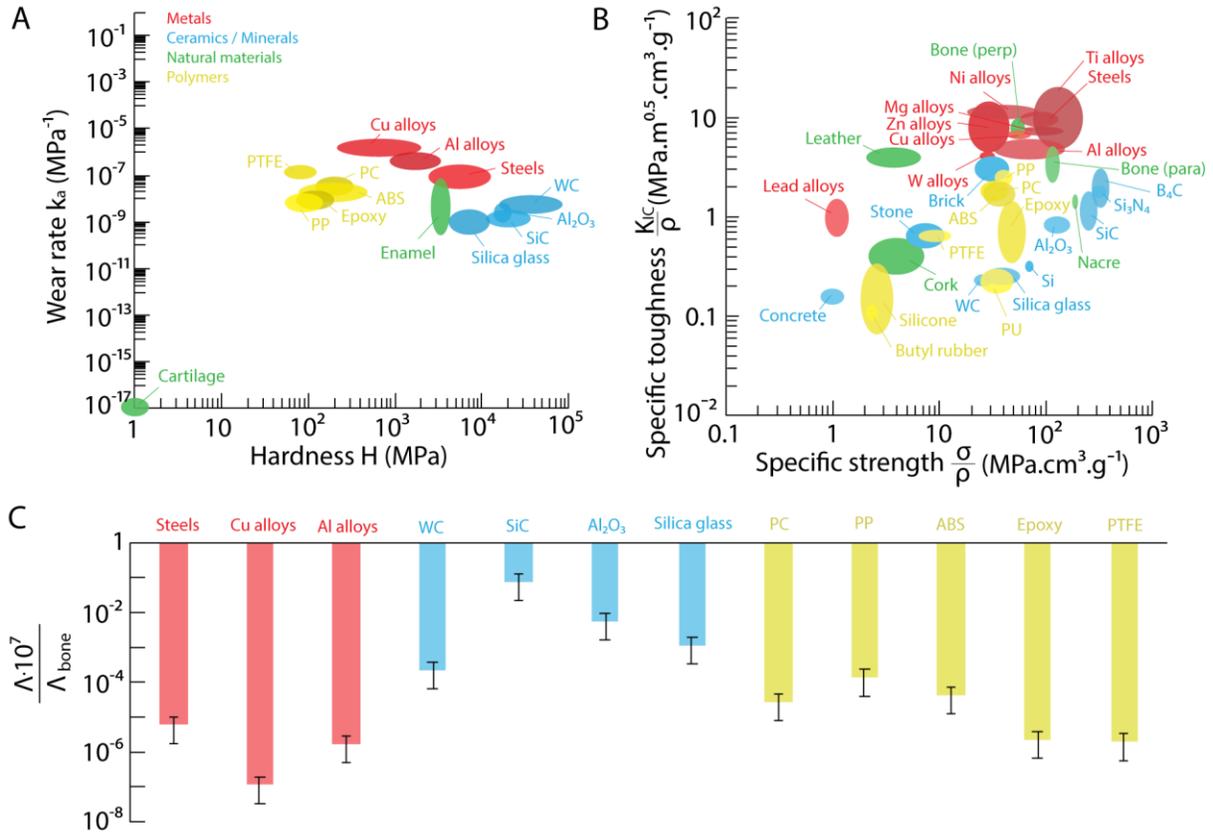

**Figure 5: Comparison of the mechanical properties of bulk materials.** A) Wear rate $k_a$ as a function of the hardness $H$, B) specific toughness $\frac{K_{IC}}{\rho}$ as a function of the specific strength $\frac{\sigma}{\rho}$, where $K_{IC}$ is the fracture toughness, $\rho$ the density and $\sigma$ the strength, and C) the comparison parameter $\Lambda$ normalized to the bone $\Lambda_{bone}$, for a selection of metals (red), ceramics and minerals (blue), polymers (yellow) and natural materials (green).

The materials to select for an implant should ideally present high wear-resistance, hardness, strength and toughness. In **Figure 5A** and **B** are materials maps (Ashby plots) that represent the wear rate as a function of the hardness and the specific toughness as a function of the specific strength for the five materials families, namely metals, composites, ceramics, polymers and biomaterials. To allow a more direct comparison of these materials properties to those of bones, a new parameter $\Lambda = \frac{H \cdot K_{IC} \cdot \sigma}{k_a \cdot \rho}$ is introduced. Normalizing $\Lambda$ with $\Lambda_{bone}$, it appears that all artificial materials, when considered as bulk without microstructuring, are quite away from the target bone-like properties (**Figure 5C**). Of the set of materials considered in this review (representative set), SiC and $Al_2O_3$ stand out as the "least worse". This indicates that ceramics in general, have more potential for bone-like mechanics, as anticipated from the high mineral content in bones (65%). A more exhaustive list of materials to compare as well as their combinations into composites could be a window for material design and selection through computational methods [47–49].

Beyond mechanics, other properties should be considered for the selection, mostly low weight and biocompatibility or bioactivity. This rules out some metals that are both heavy and prone to oxidation and corrosion [50]. A chemistry close to the natural bone is the golden dream for implants. Calcium phosphate particles of various crystallinity, shapes and dimensions have been obtained by sol-gel, precipitation, hydrothermal or microemulsion methods [51] and used for bioceramics and biocomposites [52]. But despite the control over the synthesis, calcium phosphate ceramics are less convenient than traditional technical ceramics because nanoparticles of controlled morphology and crystallinity are not easily obtained in large quantities [53], suspensions have a tendency to agglomerate [53], and the crystallographic structure changes during the sintering process used for densification and is usually associated with an increase in volume creating internal tension and cracks [54,55] (see Table 6 in the ESM for the crystallinity).

### 2.2. Material organization

Bone has a hierarchical structure with elongated and cylindrical microstructures, the osteons. Most current implants, on the contrary, have a homogeneous microstructure and rely on macroscale organization by combining several parts. Mimicking the microstructural complexity to achieve more closely the mechanics of bone is discussed in the following using nacre as an example of material where microstructure only allows the combination of strength with toughness.

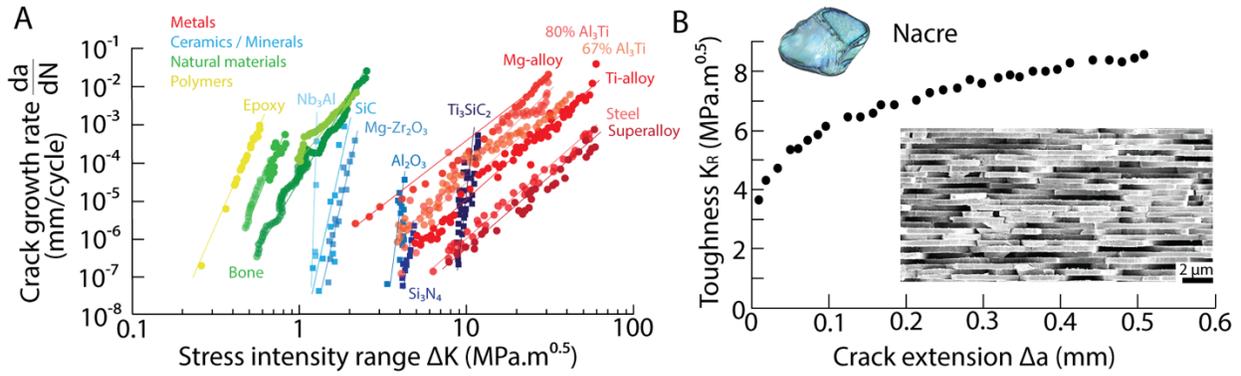

**Figure 6: Crack propagation resistance of materials.** A) Crack growth rate as a function of the stress intensity range for a selection of metals (red), ceramics (blue), polymers (yellow) and bone (green) plotted in a log-log scale. The crack growth rate is given by the derivative da/dN and is obtained by measuring the crack length during a fatigue test. B) Resistance curve (R-curve) of the nacreous layer of seashell (image, top) showing the toughness $K_R$ as a function of the crack extension $\Delta a$ (data points extracted from [56]). The R-curve implies that an increasing amount of energy is required to propagate a crack further. The electron micrograph is a view of the cross-section of nacre showing the brick-and-mortar assembly of $CaCO_3$ microtablets (image courtesy of Tobias Niebel).

The graph in **Figure 5C** shows that ceramic materials like SiC and $Al_2O_3$ are promising for use in bone implants. However, ceramics remain brittle, *i.e.* they exhibit low fracture toughness ($K_{IC}$). In bones, $K_{IC}$ ranges between 1.8 and 5 $MPa\sqrt{m}$ in the perpendicular direction to its length, and 6 to 9 in the parallel direction [57,58]. In addition, and opposite to most ceramics, bones do not break in a catastrophic manner. In traditional monolithic ceramics (with the exception of Zirconia [59]), as a crack propagates, it consumes energy to create two new surfaces. In bones, the osteon structures act as crack deflectors to absorb energy due to the increased creation of open surfaces. A second dissipative energy mechanism that occurs in bones is through soft organic layers. These are two extrinsic toughness mechanisms that directly result from the microstructure [58,60]. On the contrary, technical ceramics have homogeneous grain sizes and composition, exempt of mechanisms for crack deflection and energy absorption. The crack growth rate in ceramic materials is thus very high in comparison to bone or other tough materials like metals, as depicted by the very steep blue curves in **Figure 6A**.

Unravelling toughening mechanisms in ceramics and ceramic composites is challenging and an active field of study [61–65]. A way to incorporate toughening mechanics in ceramics is to create weak extended interfaces where cracks can be deflected and dissipate energy by the opening of new surfaces. These interfaces can be made artificially by laser cutting [66,67] or built within the material's microstructure by using anisotropic grains dispersed in a matrix [68,69]. The most wide-spread toughened ceramics are nacre-like ceramics that reproduce the brick-and-mortar microstructure of seashells (**Figure 6B**) [70,71] that exhibit a rising resistance curve (R-curve) from 3 to 9 $MPa\sqrt{m}$ (**Figure 6C**) [56]. This R-curve implies that an increasing amount of energy is required to propagate a crack further, corresponding to an increase in the toughness, noted $K_R$ in the R-curves. Nacre has actually been used as implant material due its biocompatibility, hardness (4 GPa) and wear resistance [72,73]. Indeed, the comparison parameter to bone for natural and hydrated nacre is $\frac{\Lambda}{\Lambda_{bone}} \cdot 10^7 = 0.048 \pm 0.006$ [74]. In the past, nacre has been used by the Mayas as tooth substitution [75]. More recently, *in vivo* studies of nacreous implants demonstrated a good biocompatibility and bonding to the bone after 10 months [76,77].

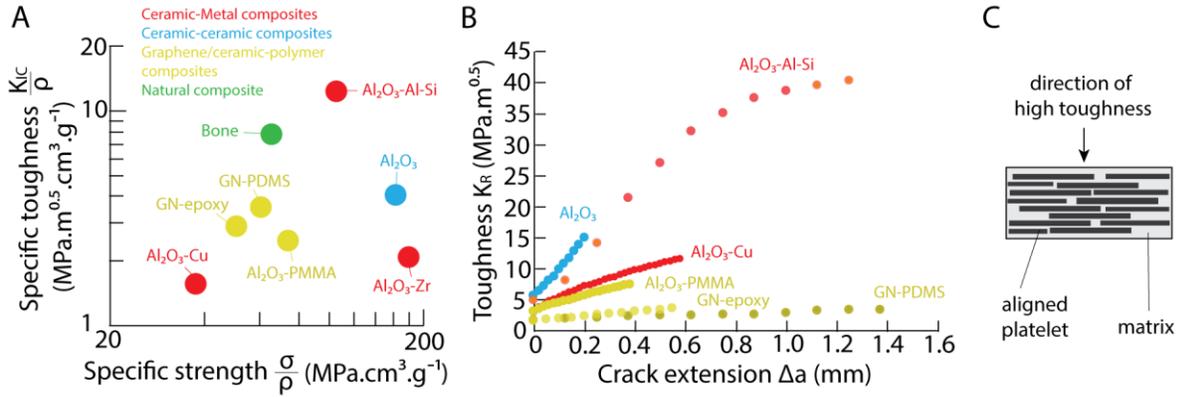

**Figure 7: Toughness of nacre-like synthetic materials and limitations.** A) Specific toughness as a function of the specific strength for selected nacre-like composites and bone (green): ceramic-metal composites (red), ceramic-ceramic composites (blue), graphene and ceramic-polymer composites (yellow). Data point are extracted from ref. [14,78–81]. B) Rising R-curves for the selected composites. Data points extracted from ref. [14,78–80]. C) Schematics of the cross-section of nacre-like composites fabricated, highlighting the unique direction of toughness and the geometrical shape.

To reproduce the properties of nacre in artificial systems, nacre-like composites have been fabricated to feature similar brick-and-mortar organization. As a result, these composites also exhibited a combination of strength and toughness with rising R-curves (**Figure 7A,B**) [14,78–82]. Although ceramic-metal composites are the most promising, the metallic mortar may pose the problem of ion leaching, corrosion and biocompatibility. Furthermore, the usage of nacre-like composites for bone implants is still hindered by their fabrication process that still lack the microstructural complexity is 3D shapes as observed in bones. Typically, their fabrication involves techniques that orient plate-like particles: external fields [78,80,83], sedimentation [84] or freeze-drying [85,86]. After orientation, the aligned specimens are concentrated by pressing at room or high temperature before infiltration with a polymer or a metal. The resulting composites exhibit only one orientation of alignment, failing to recreate the curved and heterogeneously distributed stress lines as depicted in figure 4D (**Figure 7C**). Without the pressing step, organic-inorganic nacre-like composites with wide range of chemistries have been fabricated but with a mineral content much lower than the 65% of bone, therefore those composites are much softer [87]. There is thus a need for new compositions and manufacturing methods to realize bulk biocompatible or bioactive composites reinforced in multiple directions.

### 2.2. Limitations of tissue engineering approaches

Another approach to bone repair is tissue engineering. In tissue engineering, a porous scaffold is designed to direct the fate of cells while providing structural support and integrity [88]. The scaffold is implanted at the side of defect and colonized by cells that take up the process of bone growth and remodeling while the scaffold degrades [89]. Without the need for a dense hard and tough material as scaffold, tissue engineering allows a lot more flexibility in complexity and design and does not suffer from the limitations discussed above. However, tissue engineering does not provide a straight-from-the-shelf solution to fracture injuries and requires a long patient immobilization, rehabilitation, and the use of drugs such as growth factors. Scaffolds that are pre-seeded with stem cells may accelerate the recovery but also may raise ethical issues and the long-term effects remain unknown.

With these considerations, structural implants that are readily available and functional for long-term bone function restoration are still desired. To this aim, selection of appropriate materials and their organization in hierarchical microstructures are a prerequisite for damage resistance and high strength, as evidenced in nacre-like composites. Reaching structural diversity and heterogeneity to reproduce the stress curves, local properties and 3D shapes of bone now demands suitable manufacturing methods.

## 3. The promise of *de novo* additive manufacturing practices

In the following are described additive manufacturing practices that have the potential for more biomimetic (or bone-like) implants. These implants are microstructured to attempt to reproduce the stress lines, the local variations in composition and the local and global mechanical properties, in 3D shapes (**Figure 8**). All those processes have at their initial stage particles suspended in a fluid. Indeed, liquid-based methods are so far the most convenient to achieve a control over microstructure and composition.

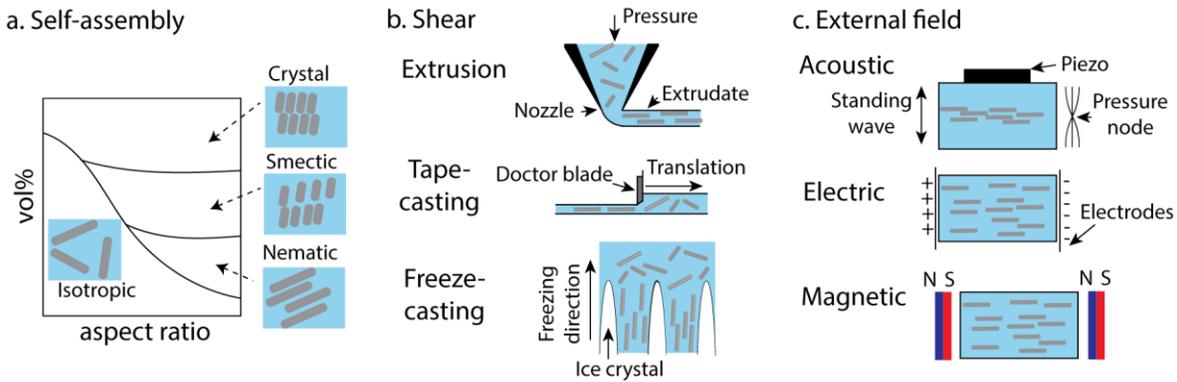

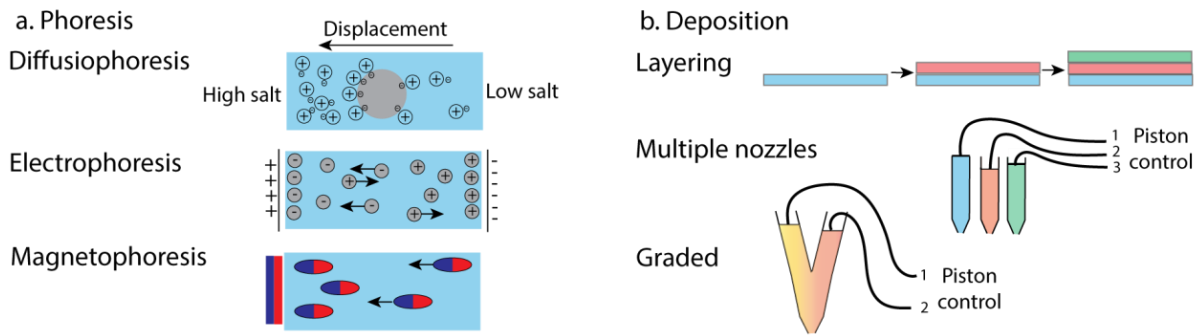

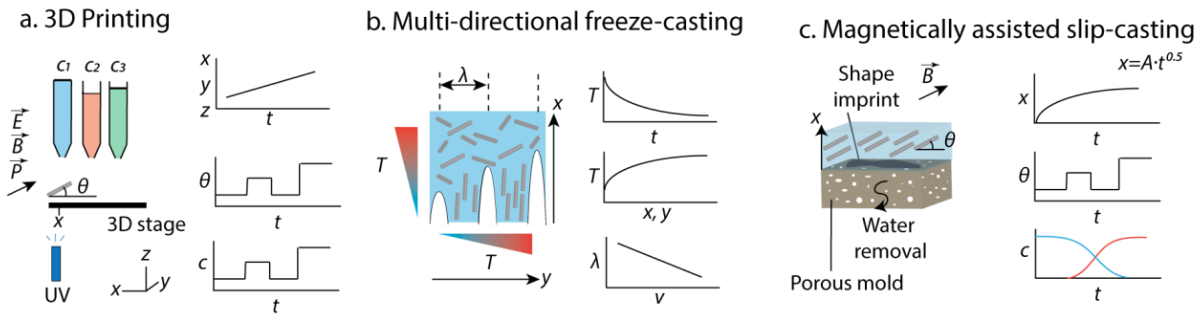

**Figure 8: Overview of advanced manufacturing practices for bone-like properties *via* biomimetic (bone-like) microstructures and properties.** (**A**) Tools to control the orientation of anisotropic reinforcing microparticles: entropically-driven self-assembly (a), shear-induced alignment (b) and external fields (c). (**B**) Methods to control the local composition: phoresis (a) and controlled deposition (b). (**C**) Methods to combine local orientation and composition controls with complex geometry: 3D printing (a), freeze-casting (b) and slip-casting (c). *x,y,z* are the three

directions of space, *t* the time, *T* the temperature, *θ* the particle angle, *c* the concentration, *λ* the structural wavelength, *v* the growth rate of ice crystals, *A* the kinetic parameter in slip-casting and $\vec{E}$, $\vec{B}$ and $\vec{P}$ the electric, magnetic and pressure field, respectively.

### 3.1. Reproducing the stress lines via alignment of reinforcement

To reproduce the stress curves of bones, there is a need to build locally aligned microstructures [30]. This can be achieved using anisotropic microrods or microplatelets oriented by self-assembly methods, shear, or external fields (**Figure 8A**). Self-assembly occurs in liquid suspensions when the concentration in Brownian particles increases and steric interactions become significant [90,91] (**Figure 8Aa**). The higher the anisotropy of the particles, the lower the concentration at which self-assembly is observed. Depending on the particle shape and particle-particle interactions, nematic, smectic, cholesteric phases and others can form. Shear is another mean to effectively induce unidirectional alignment of anisotropic nano, micro and millimetric particles (**Figure 8Ab**) [92]. Common examples of shear-induced alignment of 1D particles like fibers are during extrusion at 3D printing nozzles and during freeze-casting where the growth of anisotropic ice crystals create shear forces [93,94]. For 2D plate-like particles, tape-casting has been largely used to orient them parallel to a horizontal surface [95]. Finally, external acoustic, electric and magnetic fields are also extremely convenient to align 1D or 2D particles of any dimensions and in any deliberate direction in space [96–98] (**Figure 8Ac**).

### 3.2. Reproducing the local properties *via* controlled composition and density

In addition to local orientation, local properties can also result from local composition. The two main ways to control the composition are either to manipulate particles themselves in the colloidal liquid, or to deposit suspensions of different compositions sequentially and locally (**Figure 8B**). The displacement of nano- and microparticles suspended in a liquid can be driven by entropic and electrostatic interactions when surface charges develop, for example at specific pH or in presence of ions or charged macromolecules (**Figure 8Ba**). In such cases, it is the local salt concentration around the particles that drives its self-displacement in the fluid, a process called diffusiophoresis [99]. Alternatively, the motion of the particles can be controlled by the application of electric or magnetic fields if the particle presents a permanent charge or a magnetic dipole [100,101]. This provides the ability to position particles at specific locations through the design of complex electrodes or magnetic fields, using for example virtual magnetic molds [101].

Local deposition of suspensions of various compositions is another straight forward means to vary the composition, for example by sequential casting, layer-by-layer deposition or multi-material 3D printing [102–104] (**Figure 8Bb**). Here, maintaining a strong bonding between the layers is crucial to avoid delamination at the interface during the densification process or the mechanical loading. To prevent this, deposition of nearly continuous gradients can help, for example by *via* mixing nozzles that co-extrude or co-deposit suspensions of different yet compatible compositions [105,106].

### 3.3. Reproducing the global properties via structural hierarchy in macroscopic shapes

Reproducing the local organization, composition and properties should be accompanied by the control over the macroscopic shape. This can be done by combining the strategies presented in earlier paragraphs with additive or advanced manufacturing methods (**Figure 8C**). 3D printing, multi-directional freeze-casting and magnetically assisted slip-casting are liquid-based methods that yield, after sintering, pressing and infiltration, to high strength composites [78,85,86,107,108]. Examples of these processes and described in another review [109].

The local control and tunability in those additive or advanced manufacturing processes is achieved thanks to the time dependence of a series of parameters. In the case of 3D printing, the choice of the composition, orientation of the particles as well as the position of the stage are timely controlled (**Figure 8Ca**). These three parameters (composition, orientation and position) can thus be varied during the entire process, while the deposited liquid is solidified by curing, for example using a UV lamp and a UV-responsive resin [102,108]. In multi-directional freeze-casting, it is the local temperature and thus the ice formation that is dependent on time and position (**Figure 8Cb**) [110–113]. The growth of the ice is acting on shearing and pushing particles into aligned structures. Variations in the solidification velocity $v$ directly impact the structural wavelength $\lambda$ in the (x,y) plane and allows for changing of the composition in the casting direction (z) [114]. Finally, in magnetically assisted slip-casting, water from the liquid suspension flows through the pores of the underlying porous mold by capillary forces (**Figure 8Cc**). This capillary action varies with time as the thickness of the deposited layer $x$ at the surface of the mold increases. During the "growth" of this deposit, the orientation of the anisotropic particles in the z-direction can be varied with time-step variations of an external magnetic fields [78,97,115]. Finally, the composition of the suspension that is casted onto the mold can also be tuned with time [78,116]. Selected examples of the resulting structures achieved with these methods are described in the following paragraph.

### 3.4. Examples of bone-like microstructures and properties

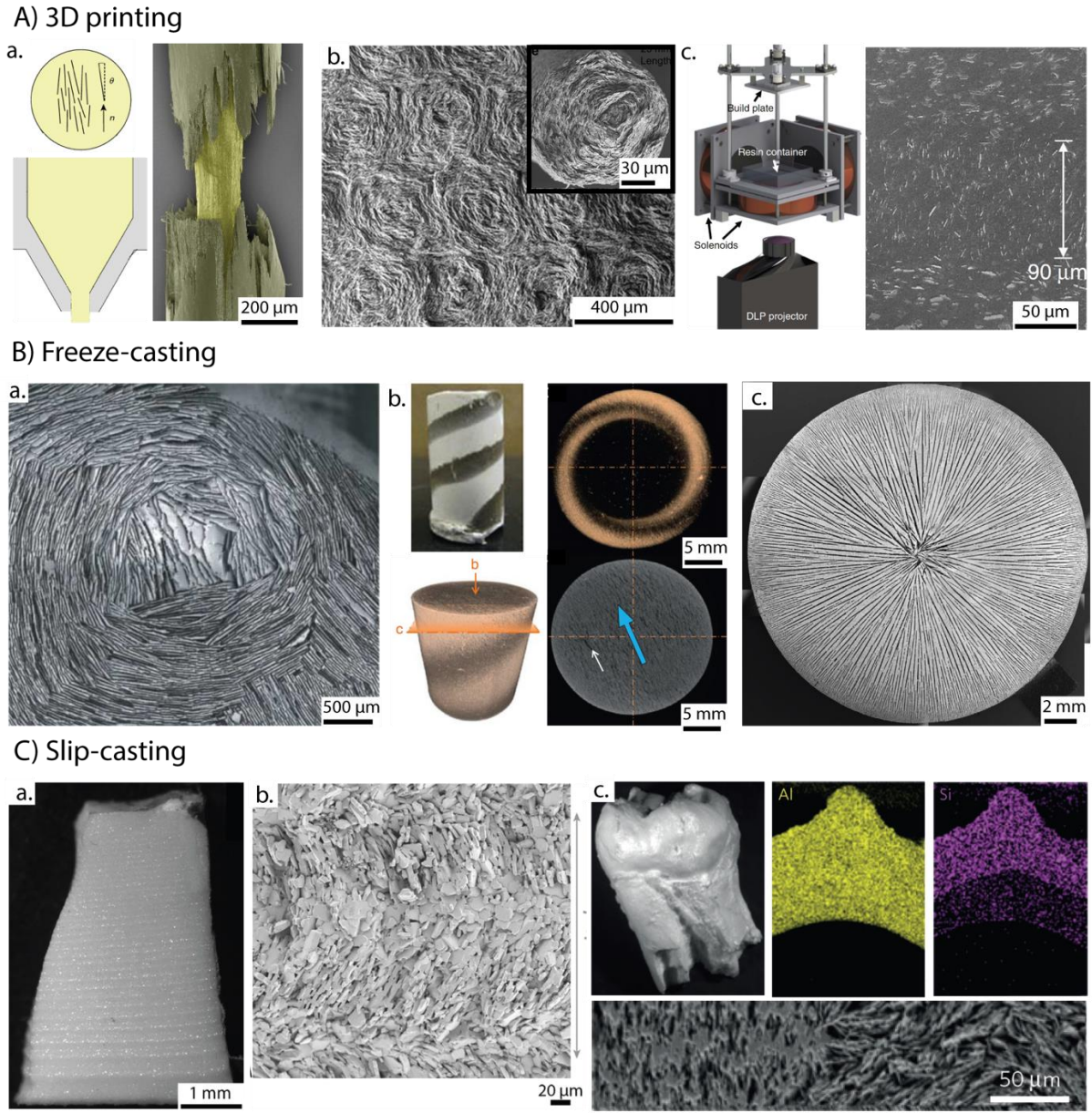

**Figure 9: Advanced and additive fabrication methods for bone-like microstructures using 3D printing (A), freeze-casting (B) and slip-casting (C). (A, a)** liquid crystal self-assembly at the extrusion nozzle (adapted from [117], copyright © 2018, Springer Nature); **(A,b)** shear-induced pattern formation using anisotropic plate-like particles (adapted from [118] copyright © 2017, Springer Nature); **(A,c)** magnetically-directed particle orientation in an UV-curing matrix (adapted from [108], copyright © 2015, Springer Nature). **(B,a)** Osteon-like pattern in freeze-

casting with controlled cooling velocity front (adapted from [119], copyright © 2006, AAAS); **(B,b)** Local orientation and composition by combination of magnetic field, freeze-casting, and sample rotation (adapted from [120], copyright © 2012, Elsevier Ltd.); **(B,c)** Concentric orientation using controlled freezing direction (adapted from [121], copyright © 2014, Elsevier Ltd.). **(C,a)** Multilayered assembly with local particle orientation using magnetically assisted slip-casting and templated grain growth (adapted from [115], copyright © 2019, The American Ceramic Society); **(C,b)** Closer view on the local grain orientation in the ceramic composite of **(C,a)** using electron microscopy (adapted from [115], copyright © 2019, The American Ceramic Society); **(C,c)** Electron micrographs of a bilayer tooth-like composite prepared by magnetically assisted slip-casting with change in orientation and composition (adapted from [78], copyright © 2015, Springer Nature).

Selected examples of bone-like complex structures obtained by additive or advanced manufacturing methods are presented in **Figure 9**. The fabricated materials exhibit high degree of control and tunability in orientation, composition, density and shape.

3D printing using direct-ink-writing (DIW) is a liquid extrusion-based method that leads to orientation of anisotropic particles at the nozzle through shear forces. Depending on the shear profile in the nozzle as well as the rheology of the ink, the microstructures obtained can show alignment of polymer nanofibrils along the printing direction that may be accompanied with a stiff skin formation [117] (**Figure 9A,a**). With concentrated suspensions of 2D microparticles, osteon-like circular patterns can also be formed (**Figure 9A,b**) [118]. After deposition, the extruded filaments retained their shape and microstructure by shear-thickening effect, or are consolidated by a curing process. To extend the controlled orientation beyond filament struts, stereolithography with external magnetic field was used to locally orient particles within a liquid matrix in an object of any shape (**Figure 9A,c**) [108]. In this case, magnetically responsive particles are suspended in a UV-curing resin and the stereolithography method is employed under a magnetic field. Basically, a mask of given geometry is placed between the uncured sample submitted to the magnetic field and a UV-lamp. This will fix the particle orientation in the area submitted to the UV, while the remaining part of the resin, stays liquid. After the curing of one layer, another one is deposited on top and the process is repeated. In direct-ink-writing, the resolution of the 3D shape is determined by the diameter of the nozzle, whereas in stereolithography, it is determined by the resolution of the mask. However, the internal resolution is determined by the particle size and orientation, namely by the nozzle and the magnetic field.

Freeze-casting can also generate osteon-like microstructures. This is realized by controlling the ice front velocity within a cylindrical mold to push anisotropic plate-like microparticles to align in concentric orientations (**Figure 9B,a**) [119]. Using mixtures of aqueous suspensions of magnetic and non-magnetic particles such as $Al_2O_3$ and $Fe_3O_4$, magnetic fields can control the position of the magnetic particles. Placing a cylindrical mold between two permanent magnets above the freeze casting set-up, and slowly spinning the two magnets during the ice formation, Porter *et al* oriented $Fe_3O_4$ particles in a helicoidal fashion [120] (**Figure 9B,b**). Instead of concentric orientations, radial orientations can also arise during freeze-casting by having two ice growth fronts perpendicular to each other (**Figure 9B,c**) [121]. The resolution of the final microstructure is thus controlled by the size of the ice crystals developed in the process as well as the response of the particles to shear and external fields, whereas the 3D shape is defined by the mold used to receive the suspension.

Finally, slip-casting with magnetic fields can make dense ceramics with controlled orientations in anisotropic 2D microparticles in a variety of pattern, such as periodic structures (**Figure 9C,a,b**) [115]. Slip-casting also allows tuning of the geometric shape and of the composition [116]. A human tooth mimic was fabricated as a proof-of-concept, with bidirectional alignment of magnetically-responsive particles of $Al_2O_3$ coated with $Fe_3O_4$ nanoparticles forming the dentin-like layer, whereas a mixture of $Al_2O_3$ and $SiO_2$ was used for the enamel-like layer (**Figure 9C,c**) [78]. The resolution of the microstructures is then controlled by the speed of the slip-casting that corresponds to the speed of water removal through the porous substrate, as well as the response of the particles to the external field and their sizes. The 3D shape resolution is however defined by porous mold.

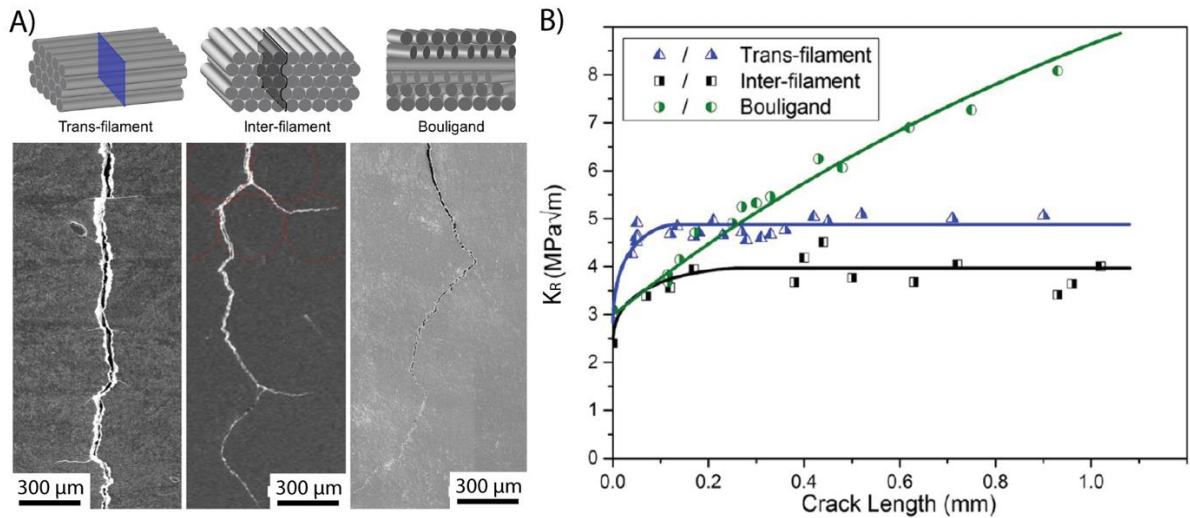

**Figure 10:** Crack propagation **(A)** and fracture toughness **(B)** as a function of the microstructure in dense and sintered $Al_2O_3$ ceramics prepared by direct-ink-writing and shear-induced orientation of anisotropic plate-like microparticles (adapted from [118], copyright © 2017, Springer Nature).

Following the presented manufacturing methods by sintering, pressing and infiltrating has been carried out to create composites and ceramics with biomimetic organization. controlled local microstructural organization that mimics that of bones. Do they achieve biomimetic properties? Generally, the local Young's modulus and hardness of these structures have been measured using nano- and micro-indentation methods, ultrasonic wave propagation, or tensile tests. Flexural strength of macroscopic beams are also commonly evaluated [85,118,122]. However, the global macroscopic mechanical properties, especially in 3D shapes generally remain unknown due to difficulties in measurement and in obtaining reproducible large specimens. One of the few examples of fracture toughness in complex hierarchical microstructures does present a rising R-curve [118] (**Figure 10**). This illustrates well how the local orientation of ceramic grains can deflect cracks and increase the damage-resistance in ceramics. In other studies, qualitative observations of crack deflection and extrinsic toughening mechanisms are made *post-mortem* [82,83,115,123,124]. For a better understanding and quantification of the role of microstructures on the mechanical properties, and to compare with those of natural bones, we need to be able to measure the mechanical properties at the macroscopic level in such complex systems. This poses multiple challenges, from the sample preparation to the models and standards to apply.

# 4. Assessing the materials performance *via* more representative characterization methods.

## 4.1. Fracture resilience in microstructured composites

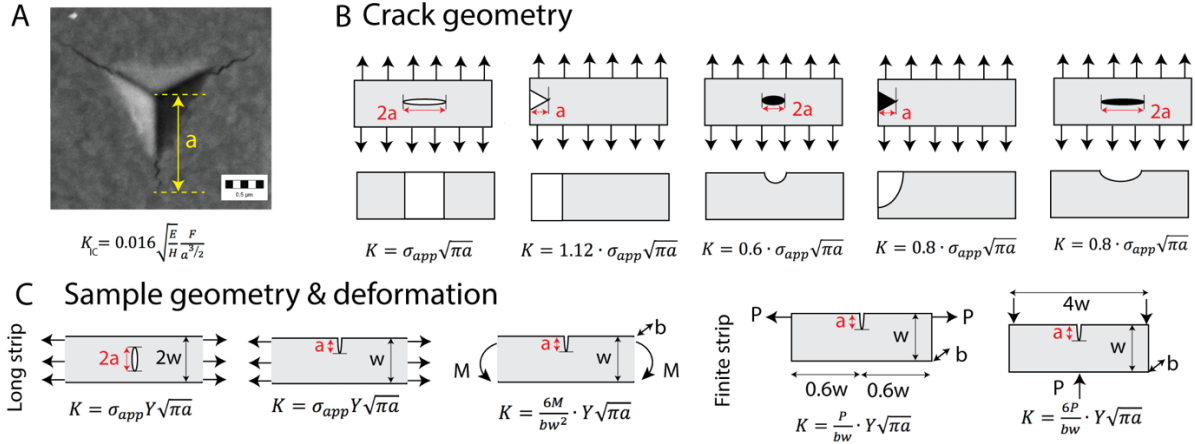

**Figure 11: Fracture toughness methods.** A) One of the common methods for the determination of the toughness $K_{IC}$ of a stiff material is nanoindentation. In this method, an indenter is punching the material under test. *Post-mortem* analysis follows to evaluate the crack length *a* emanating from the center of the indent (adapted from [124], copyright © 2015, Elsevier Ltd.), B) Possible crack geometries and corresponding stress intensity factors *K* as a function of the crack length *a* and the "far field" applied stress $\sigma_{app}$, that is indicated by the arrows. C) Possible sample geometries and deformations and the corresponding equations as a function of the crack length *a*, the applied stress $\sigma_{app}$, a parameter Y that depends on the specimen and crack geometries, the moment *M*, load *P*, width *b* and height *w*.

Fatigue fracture is the main cause of failure in bone implants at the materials level. To characterize the resistance to failure, the toughness is measured through the stress intensity factor *K* that measures the energy to input to propagate a crack (**Figure 11**). Nanoindentation has been for long a convenient *post-mortem* method to measure $K_{IC}$ based on the Lawn-Evans-Marshall model (**Figure 11A**)[125]. However, nanoindentation is known to produce inflated fracture toughness values for multiphase materials [126]. A more accurate measurement of the toughness of a material is through the strain energy release rate *J* as the crack propagates (known as J integral) [127]. (Discussion about the J integral goes beyond the focus of this paper.) In this scenario, a pre-crack is made in the sample before the loading is applied. The sharpness, size, and the geometry of the crack and of the samples, as well as the loading direction, greatly influence the output toughness value (**Figure 11B,C**).

In heterogeneous microstructured hard materials with complex geometry, this is further challenged. Fracture tests using several crack geometries and loading conditions using the traditional or novel methods [128,129] need to be performed to achieve a comprehensive overview of the materials toughness. This has been hinted in recent studies using additively manufactured bio-inspired composites [130]. However, these composites have resolutions much lower and features much larger than those discussed in this document, in the order of 50-100 µm, that is more than 10 times larger than the largest dimensions of the oriented particles as described here. The feature size and the resolution of the microstructures are important parameters as they will drive crack propagation.

## 4.2. Limits of traditional ASTM standards

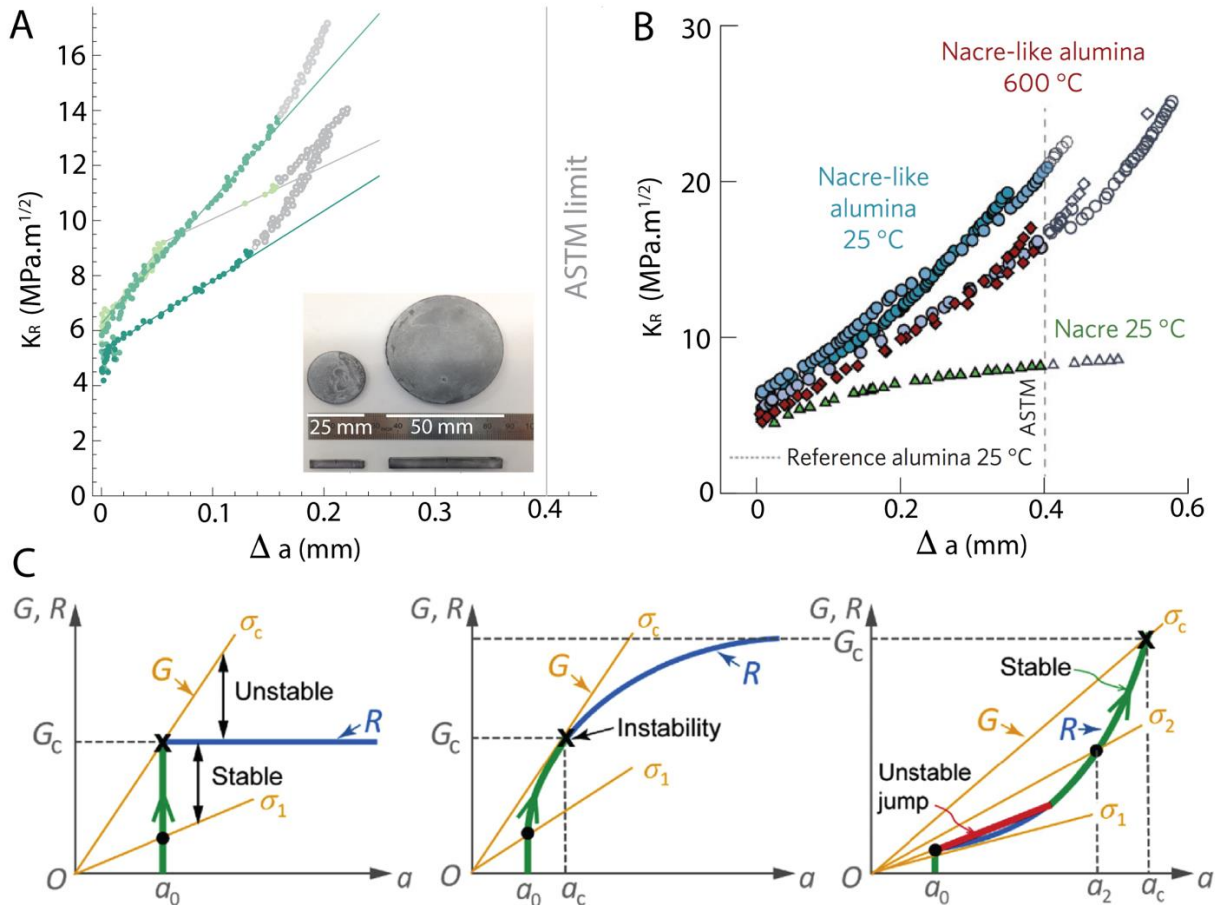

**Figure 12: Experimental observation and interpretation of rising R-curves.** R-curves of alumina nacre-like composites with 99% density for two sample thicknesses, namely 2.5 and 5 cm (**A**) and temperatures, 25 and 600 °C (**B**). The R-curves were measured using the single edge notch method with a pre-crack of length ~40 µm and in a 3-points bending set-up at a constant displacement rate of 1 µm/s. The ASTM lines in red indicate the theoretical validity limit ([122,131] copyright © 2017, Springer Nature, copyright © 2017, Cornell University). **C)** R-curve shape in the case of a brittle material (first graph, left), in the case of a microstructured composite with a fracture process zone gradually increasing then saturating (middle), and in the case of a microstructured composites with a crack initiation and a crack extension stage that remains stable until the tested specimen separates in two ([130], copyright © 2019, Acta Materialia Inc.).

In addition to the crack and sample geometries, samples with complex microstructure may also defy the traditional ASTM standard C1421 [132], *i.e,* in the case of size effects. Size effect can be defined as follows: when the crack length overpasses the uncracked ligament length in the tested beam, the R-curves adopts an upward concave shape that is not representative of the crack deflection of a virtually infinitely thick sample. To avoid this effect, the ASTM standards such as ASTM E561 set conservative limits above which the measurement is not considered valid. However, this is not always feasible. For example, in antler bone [133], the thickness of the specimen *per se* is not adequate for obtaining a reliable toughness value based on the current ASTM validity criteria.

Thanks to controllable manufacturing techniques, it is now possible to produce large samples exhibiting reproducible cracking behaviors. Yet, in those microstructured samples, the concave rising R-curve still remains, and this below these conservative limits [78,85,130] (**Figure 12A,B**). Several suggestions have been made to better understand these unusual R-curves [130,131,134]. The first trial suggestion was that the ASTM standard does not apply for heterogeneous materials. In the case of complex heterogeneous microstructured materials, local orthotropy and anisotropy have to be considered carefully. The second suggestion calls for precautions on the crack size determination. Indeed, microstructuring and heterogeneities toughen composites and ceramics through extrinsic toughening mechanisms where the propagating crack is deflected at weak interfaces. In a complex arrangement of those interfaces, the crack path becomes 3D. However, it is projected crack length that is measured and used for the determination of the R-curve. Methods such as *in situ* X-ray computed tomography might help to provide a more accurate measurement of the actual crack length, in 3D samples [135–137]. Finally, it has been suggested that some bio-inspired microstructures could be able to generate concave R-curves [130]. This is the case for the Bouligand or plywood structure and was explained by the following: in the scenario of a load-controlled dynamic impact, the J-integral exhibits a stable growth at large crack extension. In the opposite, in a displacement-controlled loading process, the crack propagates in a stable manner until reaching a threshold determined by the size effect (**Figure 12C**) [130]. These could be the first considerations to carefully look at now that processing capabilities allow large and diverse sample sizes, shapes, and microstructures, while mechanical testing can be complemented with new powerful *in situ* characterization tools such as *in situ* SEM and tomographic imaging.

### 4.3. Nature-inspired characterization methods

The characterization of complex microstructures for bone implants not only demands better experimental and theoretical understanding of crack propagation mechanisms, but also is further challenged by the various and complex types of loadings that those materials are submitted to in the body.

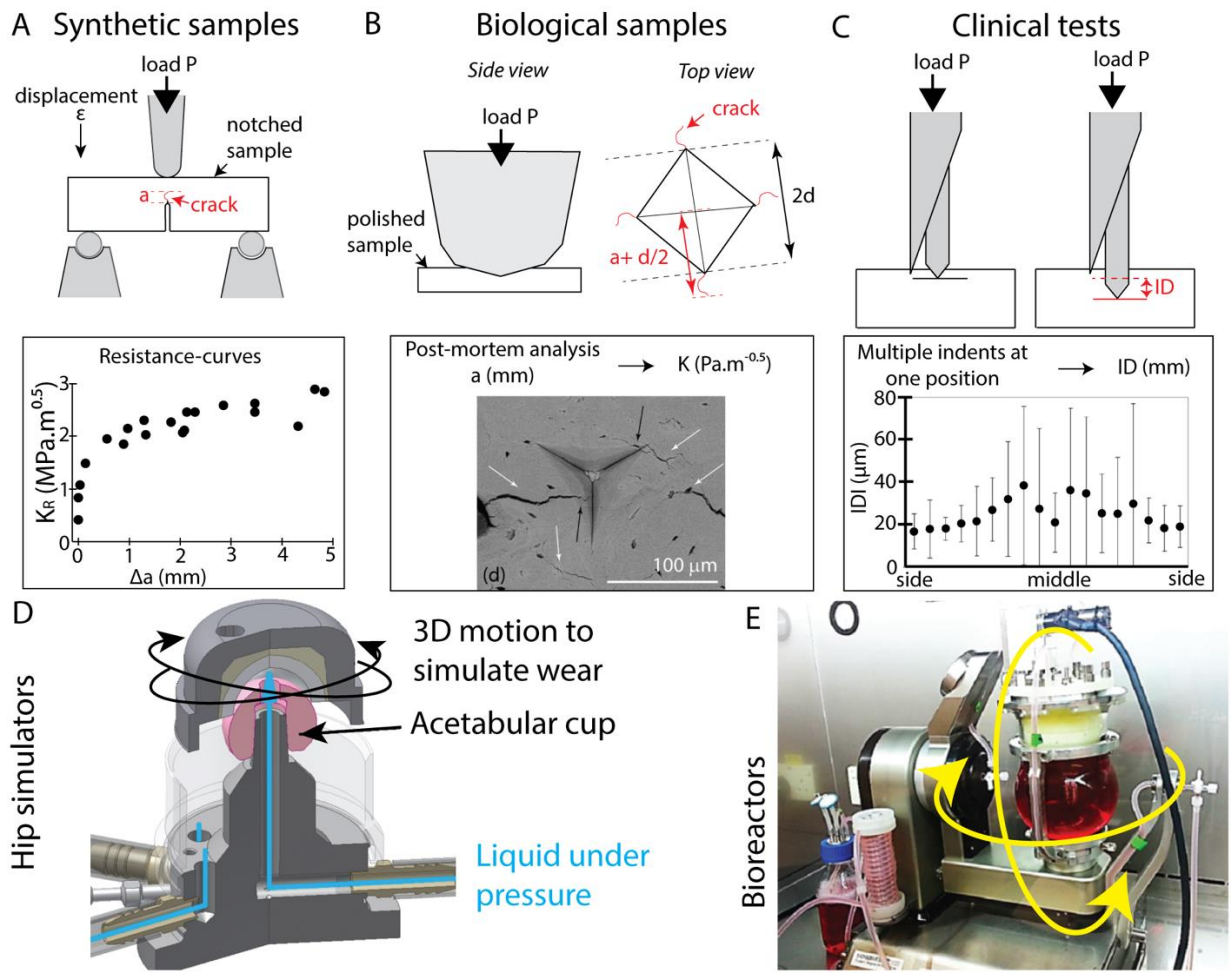

**Figure 13: Mechanical testing methods**: crack propagation resistance as measured in engineering laboratory on synthetic specimen (**A**), *post-mortem* crack assessment by nanoindentation method, as used for biological samples (**B**), and micro-indentation with reference point as used in the clinics on patients measured across the cross-section of a bone [138] (**C**), hip simulators (**D**) ([139], copyright © 2017, PLos ONE) and bioreactors (**E**) ([140], copyright © 2018, John Wiley & Sons, Ltd.). The inserts in A, B, C are examples of the results obtained using the methods on middle-aged human cortical bone: Resistance-curve, data points extracted from [141], indentation ([142], copyright © 2009, Elsevier Ltd.) and indentation distance increase across the bone from the first and the tenth loading ([138], copyright © 2015, Elsevier Ltd.)

To apply artificial materials that results from research to real clinical applications, the first challenge to tackle is the alignment of the measurements of the toughness, in particular, between the different communities, from materials science and engineering, biology and medicine, so to select the best material to be used as a bone implant. Indeed, in a mechanical lab, there is virtually no constrain in the variety of measurement configurations and sample geometries [143] (**Figure 13A**). Biological samples, on the other hand, have limitations in size, geometry, number and hence are generally characterized using nanoindentation methods that only capture the mechanical behavior locally, failing to provide a consistent overview of the mechanical properties of the composite material as would be required from the solid mechanics perspective [144–147] (**Figure 13B**). Nanoindentation is currently being complemented by *in situ* crack propagation tests in electron microscopes but still remains a local method [148]. Finally, in the clinics, the patient's bone mechanical properties are quantified using a hand-held tool that impacts the bone and measures how far the indent penetrates after several impacts [138] (**Figure 13C**). The comparison of the results obtained from those tests is thus difficult.

The second challenge is to mimic not only the material but also the complex *in vivo* loading and environmental conditions. Several joint simulators have been developed to capture the constraints that bone and joints undergo in the body [139,149] (**Figure 13D**). Some of these simulators can be operated during a large number of cycles, under various temperatures and hydration conditions. Finally, bioreactors are being developed to study the response of cells and materials under shear flow in physiological environment (**Figure 13E)** [140]**.** Such 3D, complex and varying loading conditions simulators are still to be exploited by scientists and engineers to study the fracture mechanics and resistance of bone and its potential implants, way earlier in the design and fabrication process.

## 5. Closing remarks

Bone is a complex biological material for which only limited quantitative information is available about its mechanical behavior. After decades of research on this topic, important questions still remain unanswered: how do bones deform? how do they fracture? What are the toughening mechanisms that act across length and timescales and result to its unique behavior? Obtaining further insights of these mechanisms will allow us to associate the structural functions of the various natural bones found in our body with their materials composition and microstructure. In the context of designing and fabricating artificial bones and medical implants such studies are required to reproduce more accurately the hierarchical and heterogeneous organization of natural bone. The efforts put into the development of advanced and additive manufacturing methods provide optimistic strategies in this direction with examples of strong and tough ceramic composites in a variety of shapes and with local properties. With these new materials, it is now required to update our mechanical characterization methods to better characterize these new materials. Indeed, the lack of a unified testing approach to characterize their mechanical properties is a major showstopper for a bench-to-bed translation. To fulfill their function as bone implants, these materials need to be mechanically robust and tough as the original part. New experimental frameworks for mechanical testing of inhomogeneous biological and bio-inspired materials need to be developed and coupled to multiscale computational models. Interesting ideas can be drawn from pioneering work on the development of mechanical testing frameworks of heterogenous materials, beyond the classical laminates or random mixtures, and supported with new theoretical approaches [150–152]. However, how such a framework could be adapted for the case of small scale biological materials with such a complexity as bone is still unclear. This perspective review did not considerate other functionalities such as bioactivity, osteoconduction, piezoelectricity, etc.

and focused on the mechanics viewpoint. For concrete applications as bone implants, bioinert materials with more complex functionalities, studying the mechanics would be only the first step for fast and effective long-term bone repair. They would provide design principles, manufacturing strategies and testing approaches for bone implant, that can then be applied to more biorelevant materials. The ultimate bone implant would provide the mechanical support as soon as it is implanted, then be considered by the body as part of it, to be remodeled by the metabolic process to virtually being indistinguishable with natural bone, after time.

## 6. Acknowledgments

The author acknowledges financial support from Nanyang Technological University with the Start-Up grant M4082382.050 and discussions with Dr. F. Bouville, Prof. S.-H. Teoh, and Mr. X. Liu.

## 7. Nomenclature

| | | | |
|---|---|---|---|
| $k_a$ | Wear rate | $\Delta a$ | Crack extension |
| $H$ | Hardness | $\vec{E}$ | Electric field |
| $\rho$ | Density | $\vec{B}$ | Magnetic field |
| $K$ | Stress intensity factor | $\vec{P}$ | Pressure field |
| $\sigma$ | Strength | $\theta$ | Orientation angle w.r. to horizontal |
| $\Lambda$ | Comparison parameter | $\lambda$ | Structural wavelength in the plane |
| $\frac{da}{dN}$ | Crack growth rate | $x, y, z$ | The 3 directions in space |
| $c_1$ | Concentration in composition 1 | $v$ | Growth rate of ice crystals |
| $t$ | Time | $A$ | Kinetic parameter in slip-casting |
| $a$ | Initial crack length | $E$ | Young's modulus |
| $\sigma_{app}$ | Applied stress | $P$ | Load |
| $Y$ | Parameter that depends on specimen and crack geometries | $M$ | Moment |
| | | $b$ | Width of the sample |
| $w$ | Height of the sample | $F$ | Maximum indentation load |
| $G$ | Energy release rate | $\varepsilon$ | Elongation |
| $R$ | Crack-resistance curve | $d$ | Half the diagonal of the indent imprint |
| $ID$ | Indentation depth | $K_{IC}$ | Toughness |